\documentclass[a4paper, 11pt]{article}
\usepackage[top=2cm, bottom=2.5cm, left=3cm, right=3cm]{geometry}
\usepackage{amsmath}
\usepackage{siunitx}
\usepackage{bm}
\usepackage{tikz}
\graphicspath{{figures/}}
\newcommand{\hide}[1]{}

\newcommand{\Ie}{I.\,e.}
\newcommand{\ie}{i.\,e.}

\newcommand{\eg}{e.\,g.}
\newcommand{\imag}{\mathrm{i}}
\newcommand{\euler}{\mathrm{e}}
\newcommand{\dif}{\mathrm{d}}

\newcommand{\sgn}{\mathrm{sgn}}

\newcommand{\sca}{\text{sca}}
\newcommand{\inc}{\text{inc}}
\newcommand{\tot}{\text{tot}}

\newcommand{\esca}{\bm{E}^\sca}
\newcommand{\escaff}{\bm{\mathcal{E}}^\sca}

\newcommand{\einc}{\bm{E}^\inc}
\newcommand{\eincff}{\bm{\mathcal{E}}^\inc}

\newcommand{\eint}{\bm{E}^\text{int}}
\newcommand{\etot}{\bm{E}^\tot}

\newcommand{\cnpw}{c_n^\text{pw}}
\newcommand{\pmn}{P_n^{|m|}}
\newcommand{\pimn}{\pi_n^{|m|}}
\newcommand{\taumn}{\tau_n^{|m|}}
\newcommand{\hno}{h_n^{(1)}}
\newcommand{\hnt}{h_n^{(2)}}

\newcommand{\gmntm}{g^m_{n,{\rm TM}}}
\newcommand{\gmnte}{g^m_{n,{\rm TE}}}

\newcommand{\crirel}{M}

\newcommand{\restsumNM}[2]{\sum_{n=1}^{#1}\sum_{\underset{|m|\le {#2}}{m=-n} }^n}
\newcommand{\sumNM}{\sum_{n=1}^{\infty}\sum_{m=-n}^n}
\newcommand{\nmax}{n_\text{max}}
\newcommand{\mmax}{m_\text{max}}

\begin{document}

\title{On the interference of the scattered wave and the incident wave in light scattering problems with Gaussian beams}

\author{Jonas Gienger\\
Physikalisch-Technische Bundesanstalt (PTB)\\ Abbestra\ss{}e 2--12, 10587 Berlin, Germany\\
\texttt{jonas.gienger@ptb.de} 
}
\maketitle


\begin{abstract}
 We consider the light scattering problem for a Gaussian beam and a (spherical) particle at arbitrary location.
 Within the beam cross section, the total electromagnetic field is the superposition of the 
 incident beam and the scattered wave. Using the Generalized Lorenz-Mie Theory (GLMT)
 as a vehicle to access such scattering problems, we discuss the mathematical modeling
 of this interference at short, large but finite and infinite distances from the scatterer.
 We show how to eliminate the errors that can arise from improper modeling in the most
 straight-forward manner, that is superimposing the scattered wave with the 
 closed-form expression for the Gaussian beam at a finite distance from the particle.
 GLMT uses a low order beam model ($s^1$), but using the known higher order
 models ($s^3$, $s^5$, $s^7$, \ldots) would not mitigate these errors as we discuss.
 The challenge lies in an appropriate description of the Gaussian beam at arbitrary distances from its focus, not in  its description on the scale
 of a particle (located in or near the focus) nor in the expressions for the scattered field. Hence, the solutions described here can readily be extended to light scattering frameworks other than GLMT
 and are thus also relevant for non-spherical particles.
\end{abstract}


\section{Introduction}\label{sec:intro}
\begin{figure}[th]
 \centering
 \begin{tikzpicture}
    \draw (0, 0) node[inner sep=0] {
    \includegraphics[scale=0.75]{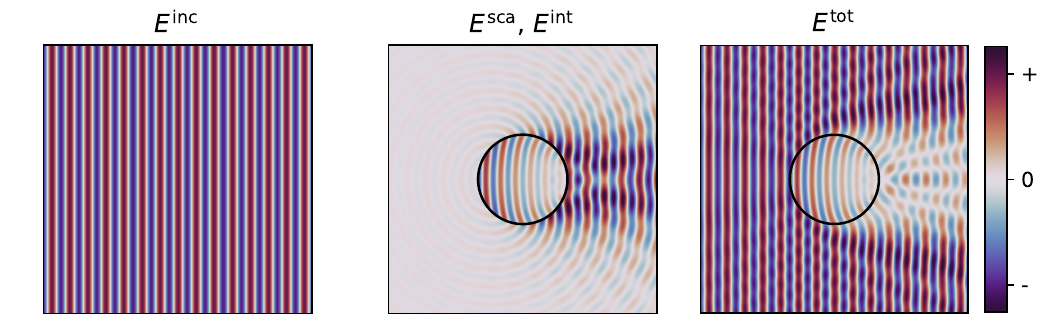}
    };
    \draw (-5.9, 1.2) node[fill=white,fill opacity=0.75,text opacity=1] {\bf\sf a)};
    \draw (-1.5, 1.2) node[fill=white,fill opacity=0.75,text opacity=1] {\bf\sf b)};
    \draw (2.45, 1.2) node[fill=white,fill opacity=0.75,text opacity=1] {\bf\sf c)};
 \end{tikzpicture}\\[5mm]
 \begin{tikzpicture}
    \draw (0, 0) node[inner sep=0] {
    \includegraphics[scale=0.75]{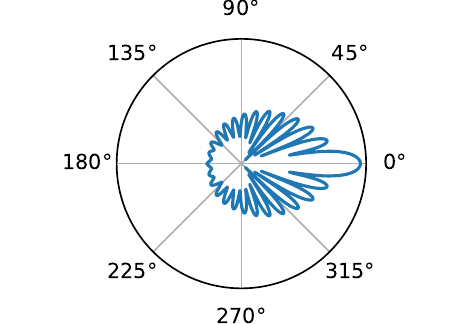}
    };
    \draw (-2.0, 1.9) node[fill=white,fill opacity=0.75,text opacity=1] {\bf\sf d)};
 \end{tikzpicture}
  \caption{
  Fields in Mie scattering exemplified for a two-dimensional scattering problem.
  a) incident field $\einc$, here: plane wave propagating from left to right; 
  b) scattered field $\esca$ (outside of particle) and internal field $\eint$ (inside particle);
  c) total field with $\etot = \einc + \esca$ (outside of particle) and $\etot=\eint$ (inside).
  d) angular distribution of far-field intensity of $\esca$ with scattering angles relative to
  incident wave.
  Panels   a) -- c) show $\Re\{E\}$ in arbitray units,  d) shows $|\mathcal{E}^\sca|^2$ (log scale), cf. Eq.~\eqref{eq:Esca_ff}.}
  \label{fig:scattering_problem}
\end{figure}

We consider the scattering of light or any other electromagnetic wave by a (small, finite) particle,
surrounded by a  homogeneous, non-absorbing host medium. For a time-harmonic field, this problem is described by the Helmholtz equation
\begin{align}
 \nabla^2 \bm{E}(\bm{r}) +  k^2 \,M^2(\bm{r})\, \bm{E}(\bm{r}) = 0
 \label{eq:Helmholtz}
\end{align}
for the Cartesian components of the  electric field $\bm{E}$. Here $k=2\pi/\lambda$ is the wavenumber with $\lambda$ the wavelength in the host medium. $M(\bm{r})$ is the (complex-valued) refractive index
relative to the host medium. Consequently, we have $M(\bm{r})=1$ outside of the particle and the particle constitutes an inhomogeneity.
For the theoretical description of scattering problems, we typically decompose $\bm{E}$ outside of the particle  into the incident field $\einc$ (field that would
be present in the absence of the particle, \eg, a plane wave or a Gaussian beam) and the scattered field $\esca$
that is caused by the presence of the particle. 
\Ie, the total field \emph{outside} of the particle is
\begin{align}
  \etot(\bm{r}) = \einc(\bm{r}) + \esca(\bm{r}). \label{eq:einc+esca}
\end{align}
This is illustrated in Fig.~\ref{fig:scattering_problem} for plane-wave scattering by a spherical particle.
Similarly, this can be written for the magnetic field $\bm{H}$, which
is, however, omitted in the following.
To solve the scattering problem, generally speaking, the ansatz in Eq.~\eqref{eq:einc+esca}
is inserted into the Helmholtz equation \eqref{eq:Helmholtz}. 
Depending on the problem at hand, one of a variety of analytical or numerical methods is then
employed to obtain a solution for $ \esca$ for a given $\einc$ and a given particle.

In many cases of practical applications which measure a portion of the field (or rather the corresponding intensity/irradiance), one is interested in the scattered field alone. This is because
the incident field $\einc$ propagates in a well-defined direction, whereas the scattered field $\esca$ generally
propagates in all directions (with varying amplitude), such that the two fields can be easily separated
at sufficient distance from the scatterer [compare Fig.~\ref{fig:scattering_problem}\,d)].
Often the incident field can be approximated
by a plane wave. An example would be a microscopic particle (\eg, a small water droplet or polymer particle) illuminated by a laser beam with sufficiently large cross section (much larger than the particle and wavelength), where we are interested in measuring how much of the laser light is scattered in a certain direction or range of directions. On the microscopic scale, such a beam has quasi-constant intensity and quasi-flat wavefronts. Hence, it
can be approximated by a plane wave on the scale of the particle. On the macroscopic scale on the other hand, such a beam has a low divergence angle, such that for many applications, we can model it as being confined to a very narrow
region around its propagation axis, which we denote as the $z$ axis. \Ie, in spherical coordinates
$(r,\vartheta,\varphi)$, $\vartheta=0$ is the \emph{forward direction}.
In such cases, considering a superposition of the 
two fields $\einc$ and $\esca$ is often only relevant for $\vartheta=0$, concerning, \eg, the phenomenon of optical extinction. \cite[chapter 3]{Bohren1998lightscatter}

There are, however, certain kinds of measurements with focused laser beams, in which the beam divergence cannot be considered to be quasi-zero and which moreover
detect the
intensity within or near the divergence angle of the beam (see end of this section for examples). Here, one has to consider both terms of the superposition $\bm{E} = \einc+\esca$
of the complex fields to model the measurement. This can lead to interesting interference effects.
In this article, we consider this problem within the framework of 
Generalized Lorenz Mie Theory (GLMT)\cite{Gouesbet1988GaussianBromwichArbitrary, Maheu1988Concise,gouesbet2017generalized}, \ie, the scattering
of a Gaussian beam (GB) by a spherical particle. We discuss how these interference effects can be described theoretically in a  consistent manner and how they can be numerically evaluated for arbitrary distances $r$ from the particle
ranging from the immediate near-field (microscopic scale) to the far-field limit (macroscopic scale)  including any intermediate values.
As it turns out, this is not as straight-forward as it may seem, given that GLMT already
provides the solution to the scattering problem.
In particular, we discuss the discrepancies occurring with an existing approach to describe the superposition
$\einc+\esca$ when $r$ is chosen too large.
While we focus on GLMT, some of the results -- mostly those regarding the behavior of Gaussian beams and the
superposition in the far-field limit -- can be applied to other light-scattering frameworks, too, if they 
include Gaussian beams, such as the Discrete Dipole Approximation (DDA) \cite{purcell1973scattering,yurkin2007DDA}.

A GB is characterized by its waist radius $w_0$, the wavelength of the light $\lambda$ or wavenumber $k=2\pi/\lambda$,
its waist location $\bm{r}_0$ and its direction of propagation\cite{Davis1979Beams,Kogelnik1966Beams,Carter1972EMfield}.
The dimensionless waist parameter 
\begin{align}
 s := \frac{1}{k\,w_0}
 \label{eq:s}
\end{align}
indicates how tightly the beam is focused, relative to the wavelength. \Ie, a small value
of $s$ indicates a wide waist and a low degree of confinement at the waist and vice versa.
The divergence angle $\theta_\text{div}$ of such a beam
is given by
\begin{align}
 \tan \theta_\text{div} = s.
\end{align}
\Ie, a tightly focused beam (small $w_0$) has high divergence and vice versa. 

\begin{figure}[t]
 \centering
 \includegraphics[width=0.6\linewidth]{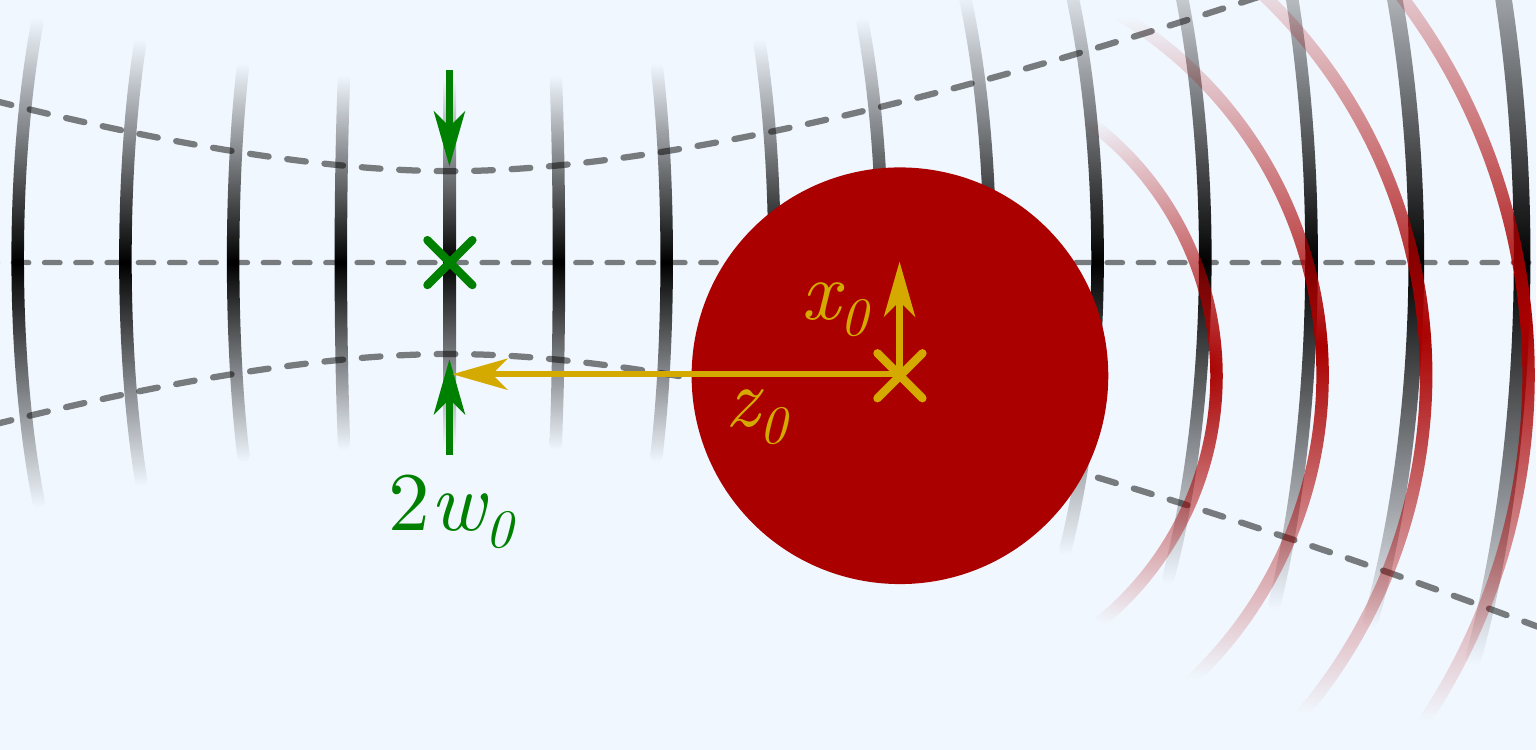}
 \caption{Schematic of the GLMT scattering problem: Spherical particle at $(x,y,z)=0$,
 $z$-propagating Gaussian beam with waist radius $w_0$ focused at $\bm{r}_0=(x_0,y_0,z_0)$.}
 \label{fig:glmt_schematic}
\end{figure}

Like a plane wave, a GB is an idealized concept
and in real-world examples its theoretical properties are not exactly fulfilled on all scales.
And while relatively simple formulas are known for the electromagnetic field of a Gaussian beam, \eg, from the classical work
of Kogelnik and Li \cite{Kogelnik1966Beams} (see section~\ref{sec:basics:GB} below), it is noted that there is no closed-form analytical expression for a
Gaussian beam that is an exact solution of Maxwell's equations.
This means that any analytical expression for a Gaussian beam
is an approximation with a certain error (which may be negligibly small in many practical applications).
Besides the lowest-order approximation \cite{Kogelnik1966Beams}, which is an approximation of the electric field of first order in the beam waist parameter $s$ (\ie, $\mathcal{O}(s^1)$, see Eq.~\eqref{eq:GB_orderL} below),
expressions of order 3, 5, 7 and 9 are known, too \cite{Davis1979Beams,Barton1989FifthOrder,Wang2005Gaussian,Luo2007Gaussian}.
The higher-order models are expected to perform better for more tightly focused beams (smaller $w_0$, \ie, larger $s$).

The considerations in this article are motivated by the interaction of microparticles with visible light
and moderately tight beam foci, for which the $\mathcal{O}(s^1)$ model still gives reasonable results.
For context, Tab.~\ref{tab:typical_parameters} provides some typical parameter values to have in mind while reading this article.
\begin{table}
 \centering
 \caption{Typical parameter values for the light scattering problems considered here}
 \label{tab:typical_parameters}
 \begin{tabular}{l r @{\;\,=\;\,} l}
   \hline
   wavelength (in the host medium)    & $\lambda$            & \SI{500}{nm} \\
   wavenumber                         & $k=2\pi/\lambda$     & \SI{12.6}{\micro m^{-1}} \\
   \hline
   particle diameter                  & $d = 2a$             & \SI{8}{\micro m} \\
   particle size parameter                     & $\alpha = k\,a$      & 50 \\
   \hline
   Gaussian beam waist radius         & $w_0$                & \SI{4}{\micro m}\\
   beam waist parameter               & $s = 1/(k\,w_0)$     & 0.02   \\
   ``lateral size parameter'' of beam &  $k\,w_0 = 1/s$                     & 50\\
   longitudinal length scale of beam  & $l = w_0/s = k\,w_0^2$              &  \SI{200}{\micro m}\\ 
   Rayleigh range                    & $z_R =l/2 = w_0^2\,\pi/\lambda$    & \SI{100}{\micro m}\\ 
     ``longitudinal size parameter'' of beam                                & $kl = 1/s^2$          & 2500 \\ 
   \hline
   distance  from particle center                         & $r$                 & $a \ldots \SI{10}{m}$\\
                                      & $kr$                 & $\alpha \ldots 10^8$
    \\ 
   \hline
 \end{tabular}
\end{table}
We are interested in the electromagnetic fields outside the particle both microscopically close to the particle and at macroscopic distances from the particle
like in a typical tabletop experiment, so ideally the expressions should work with reasonable accuracy for 
$kr  \approx 10^2 \ldots 10^8$. A formal far-field limit, \ie, $r\to \infty$ can be taken to determine the asymptotic behavior, but the 
conditions for ``$r\to \infty$'' need to be clarified.
Of course, the considerations are equally valid for scattering problems in any other part of the electromagnetic spectrum whose  dimensionless parameters $\alpha$, $s$, $kr$ etc. (see Tab.~\ref{tab:typical_parameters}) fall within a similar range.

The most straight-forward approach (called the ``naive approach'' in the following) to compute the total field for such a scattering problem would be to use (i) a closed-form expression for the GB
$\einc(\bm{r})$
and (ii) the solution for $\esca(\bm{r})$ from the appropriate (analytical or numerical) scattering solver (here GLMT) and then simply form the sum of the complex fields $\etot(\bm{r}) = 
\einc(\bm{r}) + \esca(\bm{r})$.
This naive approach is implemented in one of the computer codes
accompanying the GLMT textbook by Gouesbet and Gréhan\cite{gouesbet2017generalized}
-- an authority in the field of GLMT. 
It works fine for sufficiently small beam waist parameters $s$ and distances $r$ that are not too large. However -- for the moment ignoring computational 
limitations, \eg, due to finite machine precision -- for other parameter values it can yield results that contradict physical expectations.
This is because -- as mentioned above -- no closed-form expression for a Gaussian beam is an exact solution to Maxwell's equations and hence, 
discrepancies will show at certain parameter values, especially at larger values of $r$.
More specifically all the expressions for Gaussian beams assume that higher order terms in $s$ are small, because 
$s$ is small. For example, the  (rather complicated) equations for the ninth-order approximation (highest-order model that was published to date) are supposed to be valid as long as terms involving $s^{10}, s^{11}, $ etc. in the electric field are small and can
thus be omitted without introducing relevant errors.
However, this makes a statement about the asymptotic behavior for $s\to 0$ at any fixed point in space $\bm{r}$. In contrast for any fixed, finite $s$ (\ie, a given beam), we can in fact obtain arbitrarily large discrepancies, if the prefactors of these small-but-finite higher-order terms become sufficiently large.
This is the case for the  theoretical $r\to\infty$ limit as well as for practically relevant finite values of $r$.
A more graphic explanation of this problem goes as follows: The closed-form GB expressions (regardless of the order of approximation) stem from a paraxial approximation of the Helmholtz equation. Here, the spherical wavefronts of the GB that are expected in the far-field limit are approximated (\ie, replaced) by parabolic ones. The small phase discrepancy between spherical and parabolic wavefronts accumulates with
distance (possibly millions of wavelengths from the beam waist) and thus leads to arbitrarily large phase errors far away from the particle.
In this article we discuss this problem in detail and explain which method for the numerical computation of such interference effects is consistent with physical intuition and experiments.
This only concerns the expressions used to evaluate $\einc$, whereas $\esca$ is computed with 
standard methods. Hence the results can be expected to be compatible with various scattering solvers
and not only with the GLMT.

The theoretical results  presented here are relevant, for example, for
the so-called 
self-reference interferometric measurement principle in optical particle characterization \cite{Potenza2015complex, potenza2015drug,villa2016shape} or 
specialized optical flow cytometry techniques that do not block the illuminating laser beam from the detector \cite{kage2021multiangle}.
In both these examples, there is not yet a rigorous method to model the interference for general
particle sizes, beam waists and particle-focus distances.

\section{Theoretical background}
This section lists all the basic equations
related to GBs as well as GLMT and its far-field limit,
as far as they are relevant for this article.

\subsection{Gaussian beams}\label{sec:basics:GB}
We consider a Gaussian beam propagating along the $z$ axis and polarized in the
$xz$ plane. It is characterized by  its  waist radius $w_0$ and its waist center 
$\bm{r}_0=(x_0,y_0,z_0)^T$. We define beam-centered coordinates $\bm{q}$
as
\begin{align}
 \bm{q} = (u,v,w) &= (x,y,z) - (x_0,y_0,z_0) = \bm{r} - \bm{r}_0.
 \label{eq:def_q_coord}
\end{align}
The dimensionless waist parameter $s$ is given in Eq.~\eqref{eq:s}.
To linear order in $s$ (``order $L$ of approximation'') the Cartesian components of the electric field read \cite{gouesbet2017generalized}
\begin{align}
  \bm{E}^{\inc,\mathcal{O}L} = 
    \begin{pmatrix}
              E_u, E_v, E_w
    \end{pmatrix}
    = 
    \begin{pmatrix}
             1,
            0, 
            -\frac{2Q\,u}{l}
    \end{pmatrix}%
            E_0 \Psi_0\euler^{-\imag k\,w},
    \label{eq:GB_orderL}
\end{align}
with
\begin{align}
  \Psi_0 &= \imag Q\,\euler^{-\imag Q\frac{u^2+v^2}{w_0^2}},\label{eq:def_Psi0}\\
  Q &= \frac{1}{\imag + 2\,\frac{w}{l}}, \label{eq:def_Q}
\end{align}
where the parameter
\begin{align}
 l = w_0/s = k\,w_0^2.
\end{align}
characterizes the longitudinal extent of the focus.
At the so-called ``order $L^-$ of approximation'' one further simplifies by dropping the $w$ (or $z$) component: $E_w = 0$.

The form of the above expressions corresponds to the formulation by Davis\cite{Davis1979Beams} (in the first order), which
at the order $L^-$ can be shown to be equivalent \cite{gouesbet2017generalized} to the widely-used equations by Kogelnik and Li \cite{Kogelnik1966Beams}:
\begin{align}
 E_u = E_0\,\frac{w_0}{W}
        \exp\left[ -(u^2+v^2)\left( \frac{1}{W^2} + \frac{\imag k}{2R} \right)\right]
        \euler^{\imag \Lambda} \, \euler^{-\imag k w}
\end{align}
with
\begin{align}
 W &= w_0\sqrt{1+ \frac{4 w^2}{l^2}},\\
 R &= w\,\left(1+\frac{l^2}{4w^2}\right),\\
 \Lambda &= \arctan\frac{2w}{l}.
\end{align}

\subsection{Generalized Lorenz-Mie Theory}
In the following, we will repeat the relevant basic equations of GLMT.
A comprehensive overview 
can be found in the book by Gouesbet and Gréhan \cite{gouesbet2017generalized}.
GLMT -- at least in the strict sense -- describes the elastic scattering of a Gaussian beam by a spherical particle, cf. Fig.~\ref{fig:glmt_schematic}. 
An $\euler^{\imag \omega t}$ time dependence is assumed for all fields. 
In GLMT,
the particle is located at the origin of the coordinate system $x,y,z$. The corresponding spherical coordinates 
are $r, \vartheta,\varphi$. In these spherical coordinates, all the fields are expanded into the eigenfunctions of the 
Helmholtz operator, see Eq.~\eqref{eq:Helmholtz}.  
An equivalent formulation in vector spherical wavefunctions (as commonly used in the  T-matrix method) instead of $r$-, $\vartheta$- and $\varphi$-components also exists \cite{gouesbet2010tmatrix}, but is not the most common notation for GLMT.

The series expansion for the components of the incident field reads
\begin{align}
  E^\inc_r &=  {E_0}\sum_{n=1}^\infty \sum_{m=-n}^n k \cnpw 
    \, \gmntm \, \left[\psi_n''(kr) + \psi_n(kr) \right]\;\pmn(\cos\vartheta)\,
  \euler^{\imag m \varphi},
  \label{eq:glmt_einc_r}
  \\
  E^\inc_\vartheta &= \frac{E_0}{kr}\sum_{n=1}^\infty \sum_{m=-n}^n k \cnpw 
  \Big\{
  \gmntm\, \psi_n'(kr) \, \taumn(\cos\vartheta)  + m \, \gmnte \, \psi_n(kr)\,\pimn(\cos\vartheta)\Big\}\euler^{\imag m \varphi},
  \label{eq:glmt_einc_th}
   \\
  E^\inc_\varphi &= \frac{\imag\,E_0}{kr}\sum_{n=1}^\infty \sum_{m=-n}^n k \cnpw 
  \Big\{
  m \, \gmntm\, \psi_n'(kr) \, \pimn(\cos\vartheta)  + \gmnte \, \psi_n(kr)\,\taumn(\cos\vartheta)\Big\}\euler^{\imag m \varphi}.
   \label{eq:glmt_einc_ph}
\end{align}
For the scattered electric field this reads:
\begin{align}
   E^\sca_r &=  {-E_0}\sum_{n=1}^\infty \sum_{m=-n}^n k \cnpw 
    \, a_n\,\gmntm \, \left[\xi_n''(kr) + \xi_n(kr) \right]\;\pmn(\cos\vartheta)\,
  \euler^{\imag m \varphi},
  \label{eq:glmt_esca_r}
  \\
  E^\sca_\vartheta &= \frac{-E_0}{kr}\sum_{n=1}^\infty \sum_{m=-n}^n k \cnpw 
  \Big\{
  a_n\,\gmntm\, \xi_n'(kr) \, \taumn(\cos\vartheta)  + m \, b_n\,\gmnte \, \xi_n(kr)\,\pimn(\cos\vartheta)\Big\}\euler^{\imag m \varphi},
  \label{eq:glmt_esca_th}
   \\
  E^\sca_\varphi &= \frac{-\imag\,E_0}{kr}\sum_{n=1}^\infty \sum_{m=-n}^n k \cnpw 
  \Big\{
  m \, a_n\,\gmntm\, \xi_n'(kr) \, \pimn(\cos\vartheta)  + b_n\,\gmnte \, \xi_n(kr)\,\taumn(\cos\vartheta)\Big\}\euler^{\imag m \varphi}.
 \label{eq:glmt_esca_ph}
\end{align}
Here
\begin{align}
    \psi_n(x):= x\,j_n (x)\quad\text{and}\quad 
    \xi_n(x):= x\,\hnt (x)
  \end{align}
are the Riccati-Bessel functions corresponding to regular and outgoing waves, respectively.
$j_n, y_n$ are spherical Bessel functions and $h_n^{(1/2)}(x) = j_n(x) \pm \imag y_n(x)$ are spherical Hankel functions.
  $P_n^m$ are the associated Legendre functions and
\begin{align}
    \tau_n^m(\cos\vartheta):= \frac{\dif}{\dif \vartheta} P_n^m(\cos\vartheta)\quad\text{and}\quad 
  {}&\pi_n^m(\cos\vartheta):= \frac{P_n^m(\cos\vartheta)}{\sin\vartheta}.
\end{align}

The scattering coefficients $a_n,b_n$ of the spherical particle are those of standard (\ie, plane wave) Lorenz-Mie Theory (LMT) and are given by
\begin{align}
 a_n &=  \frac { \crirel \,\psi_n(\beta)\,\psi_n'(\alpha) - \psi_n(\alpha)\,\psi_n'(\beta) }%
				  { \crirel \,\psi_n(\beta)\,\xi_n'(\alpha) -   \xi_n(\alpha)\,\psi_n'(\beta) }
				  ,
    \label{eq:LMT_an}
    \\
 b_n &=  \frac { \psi_n(\beta)\,\psi_n'(\alpha) - \crirel \,\psi_n(\alpha)\,\psi_n'(\beta) }%
                { \psi_n(\beta)\,\xi_n'(\alpha) -  \crirel \, \xi_n(\alpha)\,\psi_n'(\beta) }
                .
    \label{eq:LMT_bn}
\end{align}
They only depend on the particle size parameter 
\[\alpha:=k\,\frac{d}{2},\] 
where $d$ is the particle diameter and 
the complex relative refractive index 
\[ M:= \frac{N_\text{particle}}{N_\text{host medium}}\]
(absorbing materials have imaginary part $\Im(M)<0$ with the sign convention used here),  $\beta = \crirel \alpha$. The above expressions Eq.~\eqref{eq:LMT_an}, \eqref{eq:LMT_bn} are for non-magnetic materials.

In contrast to the standard LMT for plane-wave scattering, which contains only
terms for $m=\pm1$, the GLMT equations contain terms for all possible indices $m=-n,\dotsc,n$. And whereas the
LMT contains only the relatively simple expansion coefficients of the plane wave
\begin{align}
 \cnpw := \frac{1}{k} (-\imag)^{n+1} \frac{2n+1}{n(n+1)}
\end{align}
the GLMT in addition contains so-called \emph{beam shape coefficients} (BSCs) $\gmntm, \gmnte$ for the focused beam. 
The BSCs for a focused beam can generally not be computed in closed form.
Besides the computation of the BSCs from the electric field by use of numerical quadratures (which is
rather computationally expensive), the BSCs can be approximated to good precision by the so-called localized approximation (LA) \cite{Gouesbet1990LocalizedInterpretation}, which is what we use here for numerical evaluation. The exact method by which the BSCs are computed should, 
however, not have any influence on the validity of the discussion in this article, because the discrepancies in the beam 
description we discuss here do not stem from the BSC computation.

\subsubsection{Far-field limit of $\esca$}
We now consider the asymptotic behavior of the scattered field in the limit $r\to\infty$, \ie, the far-field.
In this limit, the field scattered by a particle behaves like an outgoing spherical wave with a direction-dependent amplitude
\begin{align}
  \esca(r, \vartheta, \varphi) \sim 
    \, \frac{E_0}{kr}\,\euler^{-\imag k r}
    \,    \escaff(\vartheta,\varphi)
    \quad\text{at } r\to \infty,
    \label{eq:Esca_ff}
\end{align}
where the symbol $\sim$ means \emph{``is asymptotically equivalent to''}, \ie, in the specified limit (here $r\to\infty$) the left hand side of the equation behaves like the right hand side up to higher order terms (here higher orders of  $1/r$). The radial component of the far-field amplitude vanishes, \ie, $\mathcal{E}^\sca_r=0$, whereas the
$\vartheta$ and $\varphi$ components remain finite. \Ie, the far-field is a transverse wave (this 
also holds for the magnetic field $\bm{H}$). It is noted that $\escaff$ only depends on the angles $\vartheta,
\varphi$ and not on the radius $r$ anymore. This is a well-known result for scattering problems in general
and not limited to GLMT. I.e.,  this behavior is found  independently of an eigenfunction expansion in spherical coordinates.

In  GLMT (and plane-wave LMT) specifically, the  corresponding expressions for $\escaff$ are found by starting from the expressions for the field components Eq.~\eqref{eq:glmt_esca_r}--\eqref{eq:glmt_esca_ph}. Using the spherical Hankel function's asymptotic behavior at $x\to\infty$ 
 \begin{equation}
    \hnt(x) \sim \frac{1}{x}\,\imag^{n+1} \,\euler^{-\imag x}\text{ as }x\to \infty
    \label{eq:hankel2_limit}
 \end{equation}
we find  with $x=kr$ for the field components
 \begin{align}
  E^\sca_r(r,\vartheta,\varphi) &\sim 0,\\
  E^\sca_\vartheta(r,\vartheta,\varphi) &\sim \frac{E_0}{kr}\,\euler^{-\imag kr} \,\mathcal{E}^\sca_\vartheta(\vartheta,\varphi),\\
  E^\sca_\varphi(r,\vartheta,\varphi) &\sim \frac{E_0}{kr}\,\euler^{-\imag kr} \,\mathcal{E}^\sca_\varphi(\vartheta,\varphi)
 \end{align}
with (dimensionless) angle-dependent amplitudes
\begin{align}
 \mathcal{E}^\sca_\vartheta(\vartheta,\varphi) = 
    \sum_{n=1}^\infty \sum_{m=-n}^n \frac{2n+1}{n(n+1)}
  \Big\{&
  \imag\, a_n\,\gmntm \,\taumn(\cos\vartheta) 
    - m \, b_n\,\gmnte\,   \pimn(\cos\vartheta)\Big\}\euler^{\imag m \varphi},
    \label{eq:glmt_esca_ff_th} 
 \\
 \mathcal{E}^\sca_\varphi(\vartheta,\varphi) = 
 \sum_{n=1}^\infty \sum_{m=-n}^n \frac{2n+1}{n(n+1)}
  \Big\{&
  -m \,a_n\,\gmntm \, \pimn(\cos\vartheta)  
    - \imag \, b_n\,\gmnte\,   \taumn(\cos\vartheta)\Big\}\euler^{\imag m \varphi}.
    \label{eq:glmt_esca_ff_ph}
\end{align}

\subsection{Numerical evaluation}
For numerical evaluation, the infinite double sums in Eqs.~\eqref{eq:glmt_esca_r} -- \eqref{eq:glmt_esca_ph} and Eqs.~\eqref{eq:glmt_esca_ff_th}, \eqref{eq:glmt_esca_ff_ph}
are truncated according to
 \begin{align}
  \sumNM \ldots \quad \to \quad
  \restsumNM{\nmax}{\mmax} \ldots .
 \end{align}
The truncation for the $m$ summation is set to a fixed value ($\mmax=20$ is used throughout this article).
A suitable limit for $n$ is found based on the particle size parameter by Wiscombe's criterion\cite{Wiscombe1980improved}
\begin{equation}
  n_\text{max}  = \left\lfloor \alpha + 4.05\,\sqrt[3]{\alpha} \right\rfloor + 2.
  \label{eq:nmax}
\end{equation}
For example, a particle with size parameter $\alpha=50$ (Tab.~\ref{tab:typical_parameters}) will require $n_\text{max}=66$ terms
according to this criterion.

\section{Statement of the problem and proposed solutions}
After having laid out all the required components, we will now
turn again to the key problem of this article:
How to compute the superposition in Eq.~\eqref{eq:einc+esca} for arbitrary position vectors
$\bm{r}$? As mentioned before, this problem is not as straight-forward as it might seem at first glance.

\subsection{Naive approach}\label{sec:naive_approach}
The GLMT textbook by Gouesbet and Gréhan \cite{gouesbet2017generalized} comes with  computer codes
for GLMT computations,  one of which (file \texttt{diffglmt.F95})  features the interference of $\esca$ and $\einc$ in the forward direction.%
This is achieved as follows:
For a given point of observation 
$(r,\theta,\phi)$ or $(x,y,z)$,
$\esca$ is computed using the GLMT expressions for the near-field [Eqs.~\eqref{eq:glmt_esca_r}--\eqref{eq:glmt_esca_ph}], \ie, a truncated series over indices $m,n$.
$\einc$ 
is computed from the analytical expression for the Gaussian beam at order $L$ of approximation [Eq.~\eqref{eq:GB_orderL}]\footnote{The original code uses
the order $L^-$ but the results are comparable for the purposes of this article.}. The (Cartesian) components of the complex vector fields are added [Eq.~\eqref{eq:einc+esca}] and the corresponding intensity can be computed as $I(\bm{r})=|\etot(\bm{r})|^2$.
This is, of course, the most straightforward approach how to compute the superposition of the fields and can readily be applied to 
any other computational method that yields the scattered near-field $\esca(\bm{r})$ for a GB. This approach yields reasonable results for sufficiently small beam waist parameters $s$ (\ie,
sufficiently wide waist radii) and distances $r$ that are not too large. However -- for the moment ignoring computational 
limitations, \eg, due to finite machine precision -- it can yield results that contradict physical expectations at larger values of $r$.
This is not to say that the order $L$ (or $L^-$) of approximation is generally not accurate enough to describe these beams in the scattering problems at hand, but rather that one has to be aware where
a closed-form expression for the GB can be used and where not.

Numerical results illustrating the problems with this approach will follow in a later version of this manuscript.

\subsection{Solution 1: Analytical far-field limit of Gaussian beam}
In order to describe the interference of $\bm{E}^\sca$ and $\bm{E}^\inc$, we also analyze the far-field limit of the latter.
For simplicity, we take the far-field in beam-centered  coordinates $\bm{q}$ and not in the particle-centered coordinates $\bm{r}$ [see Eq.~\eqref{eq:def_q_coord}].
This means taking
the limit $q\to \infty$, where $q,\theta,\phi$ 
are the spherical coordinates
corresponding to the Cartesian coordinates $u,v,w$. 
The requirement for ``$q\to\infty$'' is 
\begin{align}
 kq\gg1,\quad  q/w_{0}\gg1, \quad q/l \gg1,
\end{align}
meaning that the radial coordinate is much larger than the wavelength, and the width and length of the beam focus.
With the additional  requirement $q/\|\bm{r}_0\|\gg1$, this limit can later be easily converted to the $r\to\infty$ limit
by multiplying with a phase factor, see Eq.~\eqref{eq:ff_phase_shift} below.

With $w=q\cos\theta$ one has in Eq.~\eqref{eq:def_Q} $Q = -\imag$ for the special case of $\cos\theta=0$, \ie, in the equatorial plane at $\theta=\pi/2$.
For the far-field, we can safely ignore this case
because for sufficiently large distances $q$ from the focus the beam intensity vanishes
away from the axis of propagation ($\cos\theta\approx \pm 1$).
Hence, we assume $\cos\theta\neq0$. 
Expanding $Q$ in powers of $1/q$ we find 
\begin{align}
 Q  =  
    \frac{l}{2 q\cos\theta}\left[ 1 - \imag \, \frac{l}{2 q\cos\theta} + \mathcal{O}\left(\frac{1}{q^2}\right) \right].
\end{align}
We denote here $\mathcal{O}(1/q^n)$ to be real-valued in order keep track of real and imaginary parts, which are each expanded to leading order separately.
Plugging this into Eq.~\eqref{eq:def_Psi0} yields
\begin{align}
  \Psi_0 \,\euler^{-\imag k\,w} = 
    \left[\frac{\imag}{k\,q\,\cos\theta} + \mathcal{O}\left(\frac{1}{q^2}\right)\right]  \, 
    &\exp\left[-\imag k q  \cos\theta \left(1+\frac{1}{2}\tan\theta^2\right) + \imag \mathcal{O}\left(\frac{1}{q}\right) \right]
    \nonumber
    \\
    \times \frac{1}{2\,s^2} & \,
    \exp\left[-\frac{\tan\theta^2}{4 s^2} + \mathcal{O}\left(\frac{1}{q^2}\right)\right]
    \label{eq:Psi0_ff_nonleading}
    .
\end{align}
Omitting any non-leading order terms in $1/q$
one finds
\begin{align}
  \Psi_0 \,\euler^{-\imag k\,w} \sim 
    \frac{\imag}{k\,q\,\cos\theta} \, 
    \euler^{-\imag k q  \cos\theta \left(1+\frac{1}{2}\tan\theta^2\right)}\,
    \frac{1}{2\,s^2}  \,
    \euler^{-\frac{\tan\theta^2}{4 s^2}}
  \quad\text{as}\quad q\to \infty
  .
  \label{eq:Psi0_ff}
\end{align}
We remind ourselves that we have taken the limit $q\to \infty$ with otherwise fixed parameters (\ie, $w_0,k,s,l$) and coordinates (\ie, $\vartheta,\varphi$). \Ie, we did not change the order of approximation in $s$.
The $\euler^{-\frac{\tan\theta^2}{4 s^2}}$ term in Eq.~\eqref{eq:Psi0_ff} will suppress the entire expression
for large $\tan\theta/s$ (or $\tan\theta/\tan\theta_\text{div}$).
This means that we can assume $\tan\theta = \mathcal{O}(s)$ and omit any terms of non-leading order in $\tan\theta$ without introducing
additional approximation errors that are not already present due to the order $L$ approximation, where $\mathcal{O}(s^2)$
terms were omitted.
In fact we \emph{must} remove those terms of non-leading order in $\tan\theta$ or $\sin\theta$, because in the 
order $L$ of approximation other terms of identical (non-leading) order were previously omitted from the equation.
Hence these
higher order terms can be regarded as artifacts from the approximate expressions for the Gaussian beam.
Leaving them in the equations for the far-field limit leads to inconsistencies and thus the occurrence of phase oscillations
without a real-world counterpart.
This happens because these terms in the complex exponentials in Eqs.~\eqref{eq:Psi0_ff_nonleading} and \eqref{eq:Psi0_ff}, while of higher order in $s$ (and hence supposedly small) have prefactors of $\mathcal{O}(q)$ which grow arbitrarily large in the far-field limit, thus leading to arbitrarily large phase errors.

Because $s\ll1$ is implied and $\sin\theta, \tan\theta = \mathcal{O}(s)$ we can approximate trigonometric functions for small 
$\theta$ in the forward hemisphere or  small $\pi-\theta$ in the backward hemisphere (paraxial beam model).
We can quite clearly see the remainder of a paraxial approximation in  the phase term in Eq.~\eqref{eq:Psi0_ff}, in the expression
$q\cos\theta \left(1+\frac{1}{2}\tan\theta^2\right)$. Going back to cylindrical beam-centered coordinates $\rho,\phi,w$ with $w = q\cos\theta$ and
$\rho = q\sin\theta$ and $q = \sqrt{w^2 + \rho^2}$, we can easily see that this is just the paraxial approximation of a spherical wavefront:
\begin{align}
 q\cos\theta \left(1+\frac{1}{2}\tan\theta^2\right) =  w\left(1+\frac{1}{2}\frac{\rho^2}{w^2}\right) =  \underbrace{w\sqrt{1+\frac{\rho^2}{w^2}}}_{= \sgn(w)\,q}{}
 + \mathcal{O}\left(\frac{\rho^4}{w^4}\right).
\end{align}
%
The difference between left-hand and right-hand side introduces a phase error that is of fourth order in $s$ (or in $\tan\theta=\rho/w$) and of first order in $q$
\begin{align}
 q - q|\cos\theta| \left(1+\frac{1}{2}\tan\theta^2\right) = q\mathcal{O}(s^4).
\end{align}
This paraxially-approximated wavefront is parabolic. 
Leaving it in the expression would cause errors and hence, now that we are in the far-field in spherical coordinates, we replace it with the spherical wavefront
that it is supposed to represent.
This leads to
 \begin{align}
    \bm{E}^\inc(q, \theta, \phi) \sim 
     \bm{\mathcal{E}}^\inc(\theta,\phi)
     \,
         \frac{E_0}{k\,q}
        \begin{cases}
    \euler^{-\imag k q}   \, &\text{for } \cos\theta > 0\\
    \euler^{+\imag k q}  &\text{for } \cos\theta < 0
   \end{cases}
    \quad\text{at } q\to \infty
     \label{eq:Einc_ff}
  \end{align}
with the vector components in spherical coordinates
\begin{align} 
  \bm{\mathcal{E}}^\inc
  =
  \begin{pmatrix}
                 \mathcal{E}_q,\mathcal{E}_\theta,\mathcal{E}_\phi
                \end{pmatrix}
   =         \frac{\imag}{2\,s^2}  \,
    \euler^{- \frac{\tan\theta^2}{4s^2}}
    \,
    \begin{pmatrix}
      0, \cos\phi, -\sgn(\cos\theta) \sin\phi
     \end{pmatrix}.
     \label{eq:Einc_ff_amplitude}
 \end{align}
From this result we see that -- just like the field scattered by the particle -- the field of the 
Gaussian beam behaves like a spherical wave with an angle-dependent envelope at sufficiently large distances from the focus. 
Because the beam itself is not radiating from any sources at finite positions (unlike the scattered field which originates from the scatterer at $\bm{r}=0$) and propagates from $w=-\infty$ to $w=+\infty$,
it behaves like an \emph{incoming} spherical wave in the backward direction ($\cos\theta<0$, $w<0$)
and like an \emph{outgoing} spherical wave in the forward direction ($\cos\theta>0$, $w>0$),
which agrees with physical intuition.

In order to transform between the particle-centered coordinates and the beam-centered  coordinates in the far-field picture, \ie,
to account for the shifted origin  $\bm{r} = \bm{q}+\bm{r}_0 $, the phase relation between $\bm{\mathcal{E}}^\inc$
and
$\bm{\mathcal{E}}^\sca$ is obtained by inserting
\begin{align}
\frac{1}{kr}\euler^{- \imag k r}
\sim  \frac{1}{kq}\,\euler^{-\imag kq} 
 \euler^{-\imag k (\bm{q}\cdot \bm{r}_0)/q} 
 \quad\text{as } q,r\to\infty \quad (q,r\gg |\bm{r}_0|)
 \label{eq:ff_phase_shift}
\end{align}
in Eq.~\eqref{eq:Esca_ff} or Eq.~\eqref{eq:Einc_ff}
and identifying $\vartheta\leftrightarrow\theta$, $\varphi\leftrightarrow\phi$.
In the forward hemisphere this allows us to compute, for example in beam-centered coordinates,
\begin{align}
    \bm{E}^\tot(q,\theta,\phi) = \bm{E}^\inc(q,\theta,\phi) + \bm{E}^\sca(r,\vartheta,\varphi) \sim 
    \underbrace{
    \left[ \bm{\mathcal{E}}^\inc(\theta,\phi) + \bm{\mathcal{E}}^\sca(\theta,\phi)\,\euler^{-\imag k (\bm{q}\cdot \bm{r}_0)/q} \right]}_{=:\bm{\mathcal{E}}^\tot(\theta,\phi)}
        \,\frac{E_0}{k\,q}\,\euler^{-\imag k q} 
    \label{eq:Etot_ff}
\end{align}
and similarly in particle-centered coordinates.

\paragraph{LMT limit}
The standard (plane-wave) LMT can be obtained as a limiting case of the GLMT by letting $w_0\to\infty$ and correspondingly $s\to0$ (from above).
Thus, we see in Eq.~\eqref{eq:Einc_ff_amplitude} that in the
LMT, the far-field behavior of $\einc$ would correspond to a $\delta$-distribution-like behavior at $\vartheta = 0,\pi$, corresponding to the exactly-defined direction of propagation of a plane wave.
Hence interference of $\esca$ and $\einc$ is not observed at any finite angles $\vartheta\neq 0,\pi$.
In GLMT however, due to the  divergence of a focused beam, $\eincff$ is nonzero for $\vartheta \neq 0,\pi$
as $r\to\infty$ 
and contains both outgoing and incoming wave components.
Hence, interference between $\einc$ and $\esca$  occurs at finite (typically small) angles $\vartheta$ (or $\pi-\vartheta$)
even in the far-field limit.

However, one has to be careful when applying this $w_0\to\infty$ limit to real-world problems, because it involves taking two limits in a row: First let $q\to\infty$ in GLMT \emph{and then}
let $w_0\to\infty$. The order is not interchangeable.
A requirement for practical applications of the far-field limit would be $q\gg l$ but after that, $l=k\,w_0^2\to\infty$ is required in the LMT limit.

\subsection{Solution 2: Compute $\bm{E}^\inc$ from the BSCs}
\subsubsection{Near-field}
We will now address the problem how to obtain a general-distance expression for the Gaussian beam $\bm{E}^\inc$ with
consistent limiting behavior for large $r$.
 Just like for the scattered field $\bm{E}^\sca$, we can compute $\bm{E}^\inc$ from series expressions in the GLMT, Eqs.~\eqref{eq:glmt_einc_r} --  \eqref{eq:glmt_einc_ph}.
 So far, these equations were a formal series expansion that was used to obtain expressions for
 $\esca$. Only the latter are evaluated numerically.
 \Ie, normally in GLMT, Eqs.~\eqref{eq:glmt_einc_r} --  \eqref{eq:glmt_einc_ph}
 serve a purely theoretical purpose and are not evaluated numerically.
 However, we can actually use Eqs.~\eqref{eq:glmt_einc_r} --  \eqref{eq:glmt_einc_ph}
 to obtain $\einc$ without the above-mentioned problems. One reason 
 for this is that -- unlike the closed-form (but approximate) GB-expressions -- 
 the series expressions are exact solutions of Maxwell's equations (but not strictly Gaussian).
 For numerical evaluation, we need to use truncated sums
 \begin{align}
  \sumNM \ldots \quad \to \quad
  \restsumNM{\nmax^\text{beam}}{\mmax} \ldots .
 \end{align}
It is noted that we are using a different truncation index $\nmax^{\text{beam}}$ when using the
truncated series for $\einc$
than for the truncated series 
for $\esca$. This is because for $\einc$ one cannot rely on the scattering coefficients $a_n, b_n$
to suppress terms in the sum above an $\nmax$ that is based on particle size (compare 
Eqs.~\eqref{eq:glmt_einc_r} --  \eqref{eq:glmt_einc_ph} with 
Eqs.~\eqref{eq:glmt_esca_r} --  \eqref{eq:glmt_esca_ph}). Instead the truncation
needs to be chosen based on the beam parameters.
In other words, it does not suffice to use those modes in the infinite series for which the
particle has a significant scattering amplitude, but we need to use all the modes that
have relevant amplitude for the beam.
In some preliminary numerical experiments, it was found that a truncation at 
 \begin{equation}
  n_\text{max}^\text{beam} = \left\lfloor k \left( 2.5\,w_0 + \sqrt{x_0^2 + y_0^2}\right)   \right\rfloor + 2
  \label{eq:nmax_beam}
 \end{equation}
 seems to work well, compare Eq.~\eqref{eq:nmax}. 
For example,  a beam with a waist radius of $w_0=\SI{4}{\micro m}$ at $\lambda=\SI{500}{nm}$ (or $k\, w_0=50$, see Tab.~\ref{tab:typical_parameters}) requires $n_\text{max}^\text{beam}=127$ terms
for the on-axis case and $n_\text{max}^\text{beam}=177$ terms for $x_0=w_0$, $y_0=0$ according to this criterion.
In contrast, for the scattered field from a particle with radius $a=\SI{4}{\micro m}$ ($k\,a=50$)
only $n_\text{max}=66$ terms are required.
However, this criterion was only tested for a limited range of parameters and its validity was not yet analyzed systematically.
 In particular it was not tested for large $z_0$, only for $z_0$ equal to zero or on the order of $w_0$.

Compared to the naive approach (Section~\ref{sec:naive_approach})
this ``$\einc$ from BSCs'' approach has the following advantages:
\begin{itemize}
  \item It is always a solution to Maxwell's equations, even with truncated series, because each eigenfunction in the series
  is Maxwellian. \Ie, the Gaussian beam is \emph{remodelled} to be strictly Maxwellian (but not strictly Gaussian anymore) \cite{gouesbet2017generalized}
  \item It is consistent with the scattered field because we are using the same field that was actually scattered by the particle.
  \Ie, even if Maxwellian beam remodelling or inaccuracies of the BSC computation have a significant influence on the scattering problem, we are at least describing this scattering problem in a consistent manner.   
\end{itemize}
Compared to Solution 1 (analytical far-field limit of GB), the advantages are:
\begin{itemize}
  \item This approach works for any finite $r$ (small or large) without issues. 
  Far-field expressions can be obtained, too (see below).
 \end{itemize}
A disadvantage of this approach, compared to using the simple analytical GB expressions in the near-field or far-field (naive approach  and  Solution 1) is that one needs to evaluate 
a (truncated) series with many terms, which increases computational cost significantly. And while a similar series expression 
needs to be evaluated in the GLMT computation of $\esca$ (which would just mean a factor of two in computational cost when $\einc$ is evaluated, too)  
the number of terms required to evaluate $\einc$ in GLMT depends on the beam geometry
(waist diameter and location) instead of particle size. In the typical case where the
beam waist is larger than the particle diameter, this can require significantly more
terms to compute $\einc$ than for $\bm{E}^\sca$. This number is typically even larger 
for off-axis beams [see Eq.~\eqref{eq:nmax_beam}].

\subsubsection{Far-field behavior}
We will now consider the far-field limit of $\einc$ in Solution 2.
 $\bm{E}^\sca$ in Eqs.~\eqref{eq:glmt_esca_r} -- \eqref{eq:glmt_esca_ph} contains Riccati-Bessel functions $\xi_n(x) = x\,\hnt (x) = x\left[j_n(x) - \imag y_n(x)\right]$.
 The asymptotic behavior is
   $\hnt(x) \sim \frac{1}{x}\,\imag^{n+1} \,\euler^{-\imag x}$ as $x\to \infty$,
 which leads to Eqs.~\eqref{eq:glmt_esca_ff_th}, \eqref{eq:glmt_esca_ff_ph}.   
 $\bm{E}^\inc$ in Eqs.~\eqref{eq:glmt_einc_r} -- \eqref{eq:glmt_einc_ph} on the other hand contains $\psi_n(x) = x j_n(x)$. In contrast to the spherical Hankel functions $\hnt, \hno$, the spherical Bessel functions $j_n(x)$ do not tend to a spherical wave
 as $x\to \infty$.
 However, we can rearrange as follows
 \begin{align}
  j_n(x) = \frac{1}{2}\left[j_n(x) - \imag y_n(x) + j_n(x) + \imag y_n(x)\right]
   = \frac{1}{2}\left[ \hnt(x) + \hno(x)  \right]
   \label{eq:Bessel_jn_separate}
 \end{align}
 and thus separate into two parts that do tend to spherical waves as $x\to\infty$.
Inserting the asymptotic behavior of Eq.~\eqref{eq:Bessel_jn_separate} at $x\to\infty$ into Eqs.~\eqref{eq:glmt_einc_r} -- \eqref{eq:glmt_einc_ph} we find
\begin{align}
 \bm{E}^\inc(r,\vartheta,\varphi) \sim \frac{E_0}{kr}\,\euler^{-\imag k r} \bm{\mathcal{E}}^\inc_\text{out}(\vartheta,\varphi) +  \frac{E_0}{kr}\,\euler^{+\imag k r} \bm{\mathcal{E}}^\inc_\text{in}(\vartheta,\varphi),
\end{align}
where $\bm{\mathcal{E}}^\inc_\text{out}$ and $\bm{\mathcal{E}}^\inc_\text{in}$ (given by series expressions)
are the far-field amplitudes for the outgoing and incoming spherical wave parts of $\einc$. 
This approach to separate the incident field (or ``beam'') into an incoming and an outgoing spherical wave was
already used by Lock \cite{Lock1995extinction}.

The resulting expressions for $\bm{\mathcal{E}}^\inc_\text{out}$ and $\bm{\mathcal{E}}^\inc_\text{in}$ 
are formally very similar to those for the scattered far-field amplitude $\bm{\mathcal{E}}^{\sca}$.
In fact they can be obtained from  $\bm{\mathcal{E}}^{\sca}$ by setting
the scattering coefficients $a_n, b_n$ to specific values as follows
\begin{align}
\bm{\mathcal{E}}^{\inc}_\text{out} = \bm{\mathcal{E}}^{\sca}\bigg|_{
            \scriptsize
            \begin{array}{l l}
            \\[-5ex]
            a_n = -1/2\\
            b_n = -1/2                                                  
            \end{array}
            },
    \qquad
\bm{\mathcal{E}}^{\inc}_\text{in} = \bm{\mathcal{E}}^{\sca}\bigg|_{
            \scriptsize
            \begin{array}{l l}
            \\[-5ex]
            a_n = (-1)^{n+1}/2\\
            b_n = (-1)^n/2                                                 
            \end{array}
            }
\end{align}
For computations, this is gives the possibility to recycle existing GLMT code for $\bm{\mathcal{E}}^{\sca}$ by simply inserting the required $a_n, b_n$ and using$\nmax^\text{beam}$ instead of $\nmax$ [Eq.~\eqref{eq:nmax}, Eq.~\eqref{eq:nmax_beam}].

Numerical experiments show that, as required, $\bm{\mathcal{E}}^{\inc}_\text{out}$ is directed in the forward hemisphere, \ie, it has non-zero amplitude only for $\cos\vartheta>0$  (within the approximation error of a truncated series representation and numerical precision). Correspondingly $\bm{\mathcal{E}}^{\inc}_\text{in}$ is directed in the backward hemisphere 
(non-zero for $\cos\vartheta<0$).

\subsection{Excursus: Higher-order beam models}\label{sec:theory:higher_order}
One might expect that the problems when superimposing the lowest-order (\ie, $\mathcal{O}(s^1)$) Gaussian beam model with the scattered field at larger distances could be mitigated 
using a higher-order beam model, such as the $\mathcal{O}(s^3)$ model by 
Davis  \cite{Davis1979Beams} or the $\mathcal{O}(s^5)$ model by Barton and Alexander \cite{Barton1989FifthOrder} and that,
possibly, the distance where artifacts occur will increase with the order of the model. Unfortunately, this is not the
case at all as we will discuss in more detail in a future version of this manuscript.

\section{Conclusion}
To conclude, in this article we have learned about the problems that can arise when 
mathematically modeling the superposition of the electromagnetic field scattered
by a (microscopic) particle with the incident field, for the case that the latter is a Gaussian beam (GB).
The closed-form expressions for 
a GB are intrinsically near-field descriptions and using them at large (macroscopic) distances
results in artifacts that do not correspond to the physical reality of the intended application -- laser beam scattering
by microparticles, for example. These artifacts occur in the form of deviations in the shape of the wavefronts (or phase errors) in
the paraxial approximation. Our analysis was focused on the lowest-order (in the parameter $s$)
GB models. However, an analysis of higher-order beam models reveals that -- somewhat counterintuitively --
the problem is not mitigated but becomes even more severe.

We presented different solutions for the problem, based on (1) correcting the 
paraxial artifacts in the analytical far-field limit of the GB and (2) using the 
series expressions in the GLMT not only for the calculation of $\esca$ but also for $\einc$.
In both cases, in the far-field the GB tends to a spherical wave which is outgoing (like the scattered wave) in the forward hemisphere and oncoming in the backward hemisphere.
Hence, in the forward hemisphere the resulting total field is described by a single
direction-dependent amplitude for an outgoing spherical wave. 
In the backward hemisphere, this results in standing waves.

The GLMT was used for the analyses in this article, as is represents the analytical solution for the
scattering problem under consideration. The findings in this article are, however, 
equally relevant to other methods for the solution of light scattering problems.
In particular, the discrete dipole approximation (DDA) is capable to handle GBs (or in fact any
sort of incident wave) because as an input for the computation it only requires
the incident field on a (usually cubic) grid of points covering the particle volume.
Typically (\eg, in the ADDA code \hide{\cite{ADDA}}), the higher-order closed-form expressions ($\mathcal{O}(s^5)$ in ADDA) are
used for this. As we have learned, this is fine as long as the GB expressions are only evaluated --
and thus as long as the particle is located --
in the near-field of the beam, \ie, near the beam waist.
However if the particle (and with it the computational grid) is moved far outside of the focus (most relevant case: large $z_0$ and significantly  non-zero $x_0^2+y_0^2$), the non-Maxwellian 
character of the GB model equation will show. With phase errors or even amplitude errors (for the higher-order beam models) present already in the input of the DDA computation, a correct result cannot be expected, particularly regarding the correct absolute phase of the 
scattered field.
Similar problems could be expected to  occur with the localized approximation in GLMT, which is also based on evaluating the closed-form GB model at specific locations (equatorial plane of the particle)
if the beam waist is far away from the particle.

Lastly, 
for GB scattering problems that are correctly solved by the method 
in question, the resulting scattered field $\esca$ from any light scattering solver can be superimposed with the GB field with the approaches described in this article, with the the analytical far-field
of the GB likely being both the simplest and the most practical.

\pagebreak
\subsubsection*{Acknowledgments}
The author thanks Daniel Kage and Toralf Kaiser [German Rheumatism Research 
Centre (DRFZ), Berlin Germany]
for sharing their experimental results that were the inspiration for this topic 
and thanks Thomas Wriedt (Leibniz Institute for Materials Engineering IWT, Bremen, Germany)
for helpful discussions.

\bibliography{bibliography}
\bibliographystyle{unsrt}
\end{document}